\documentclass[prb,aps,twocolumn,amsmath,amssymb,superscriptaddress,longbibliography]{revtex4-2}

\usepackage[dvips]{epsfig}
\usepackage{float}
\usepackage{color,soul}
\usepackage[sort&compress]{natbib}
\usepackage[english]{babel}
\makeatletter
\renewcommand{\@biblabel}[1]{#1. }
\renewcommand{\@dotsep}{500}
\renewcommand{\@pnumwidth}{0em}
\renewcommand{\l@figure}[2]{
\@dottedtocline{1}{1.5em}{2em}{Figure #1}{}\vspace{15pt}}

\makeatletter
\def\bbl@set@language#1{%
  \edef\languagename{%
    \ifnum\escapechar=\expandafter`\string#1\@empty
    \else\string#1\@empty\fi}%
  \@ifundefined{babel@language@alias@\languagename}{}{%
    \edef\languagename{\@nameuse{babel@language@alias@\languagename}}%
  }%
  \select@language{\languagename}%
  \expandafter\ifx\csname date\languagename\endcsname\relax\else
    \if@filesw
      \protected@write\@auxout{}{\string\select@language{\languagename}}%
      \bbl@for\bbl@tempa\BabelContentsFiles{%
        \addtocontents{\bbl@tempa}{\xstring\select@language{\languagename}}}%
      \bbl@usehooks{write}{}%
    \fi
  \fi}
\newcommand{\DeclareLanguageAlias}[2]{%
  \global\@namedef{babel@language@alias@#1}{#2}%
}
\makeatother

\DeclareLanguageAlias{en}{english}

\begin{document}
\title{100~GHz Bandwidth, 1 Volt Near-infrared Electro-optic Mach-Zehnder Modulator}
\author{Forrest Valdez}
	\email{fgvaldez@eng.ucsd.edu}
	\affiliation{University of California, San Diego, Department of Electrical and Computer Engineering, La~Jolla, California 92093-0407, USA}
\author{Viphretuo Mere}
	\email{vmere@eng.ucsd.edu}
	\affiliation{University of California, San Diego, Department of Electrical and Computer Engineering, La~Jolla, California 92093-0407, USA}
\author{Shayan Mookherjea}
	\email{smookherjea@ucsd.edu}
	\affiliation{University of California, San Diego, Department of Electrical and Computer Engineering, La~Jolla, California 92093-0407, USA}
\date{\today}

\begin{abstract}	
An integrated hybrid thin-film lithium niobate (TFLN) electro-optic Mach-Zehnder modulator (MZM) is shown at near-infrared wavelengths. The design uses TFLN bonded to planarized silicon nitride waveguide circuits, and does not require etching or patterning of TFLN. The push-pull MZM achieves a half-wave voltage length product ($V_\pi L$) of 0.8 V.cm at 784 nm. MZM devices with 0.4 cm and 0.8 cm modulation length show a broadband electro-optic response with a 3 dB bandwidth beyond 100 GHz, with the latter showing a bandwidth to half-wave voltage ratio of 100 GHz/V. 
\end{abstract}

\maketitle
Historically, electrically-driven modulation of laser light at near-infrared (NIR) wavelengths around $0.8\ \mathrm{\mu m}$ was used in compact disc technology~\cite{heemskerkCompactDiscSystem1982}, spectroscopy~\cite{barwoodFrequencyMeasurementsOptically1991} and optical communications~\cite{hackertPerformanceTelecommunicationgradeMultimode1992,giggenbachOpticalDataDownlinks2009}. These wavelengths are less absorbed by water and penetrate more deeply into skin, and are used in sensors, biological and chemical applications~\cite{allenDiodeLaserAbsorption1998,yooLowPowerOpticalTrapping2018} and phototherapy~\cite{andersonSelectivePhotothermolysisPrecise1983,yunLightDiagnosisTherapy2017}. NIR optical modulators are used to generate pulse sequences in quantum optics~\cite{schaferFastQuantumLogic2018} and for qubit readout~\cite{stasRobustMultiqubitQuantum2022}. In communications, externally-modulated lasers have not outperformed directly-modulated lasers in optical interconnects at NIR wavelengths, whereas the opposite is true in the telecom band~\cite{kuchtaHighCapacityVCSELbased2017}. Improved performance, integration, and scalability of NIR electro-optic (EO) modulators can thus benefit many applications including communications and information processing, sensing and spectroscopy, biomedical sciences, and laser technology. 

At these wavelengths, passive devices, nonlinear optical components and optomechanical modulators have been developed using many materials such as silicon nitride (SiN)~\cite{mossNewCMOScompatiblePlatforms2013,subramanianLowLossSinglemodePECVD2013a,geuzebroekPhotonicIntegratedCircuits2016,munozFoundryDevelopmentsSilicon2019}, alumina ($\text{Al}_2\text{O}_3$)~\cite{sorace-agaskarVersatileSiliconNitride2019}, aluminum nitride~\cite{dongHighspeedProgrammablePhotonic2022a}, tantala ($\text{Ta}_2\text{O}_5$)~\cite{beltUltralowlossTa2O2017,jungTantalaKerrNonlinear2021} or III-V semiconductors~\cite{wilsonIntegratedGalliumPhosphide2020}. However, high-bandwidth EO modulators are also needed, and one recently-studied approach uses etched thin-film lithium niobate (TFLN) waveguides fabricated on lithium niobate on insulator (LNOI) wafers~\cite{celikHighbandwidthCMOSvoltagelevelElectrooptic2022a,liHighModulationEfficiency2022a,renaudSub1VoltHighBandwidth2022}. Our work reported here uses TFLN but with hybrid modes in which the mode fraction of light in TFLN is controlled over a wide range by varying the width of the waveguides in the SiN layer alone. No etching or patterning of TFLN is required, and  TFLN is bonded at room temperature to chips from a planarized wafer after CMOS-compatible SiN fabrication was completed. The hybrid integration approach facilitates the back-end integration of non-CMOS compatible materials such as LN as part of a larger photonic circuit fabricated on standard silicon-on-insulator (SOI) wafers.   

While no single metric captures all aspects of EO performance, an often-used figure-of-merit for comparison is the ratio of 3-dB EO bandwidth (GHz) to the voltage ($V_\pi$) required to achieve a $\pi$ phase shift~\cite{sinatkasElectroopticModulationIntegrated2021}. Here, we report integrated EO Mach-Zehnder modulators (MZM) based on Pockels effect in 5-mol\% MgO-doped TFLN integrated with SiN waveguides, which achieves 100 GHz bandwidth (3 dB EO bandwidth) with $V_\pi$ = 1 V at 784 nm. This combination improves upon the performance of NIR integrated MZM devices, and achieves a new milestone for EO modulators in general.
\begin{figure*}[tbh]
\centering
\includegraphics[width=\linewidth, clip=true]{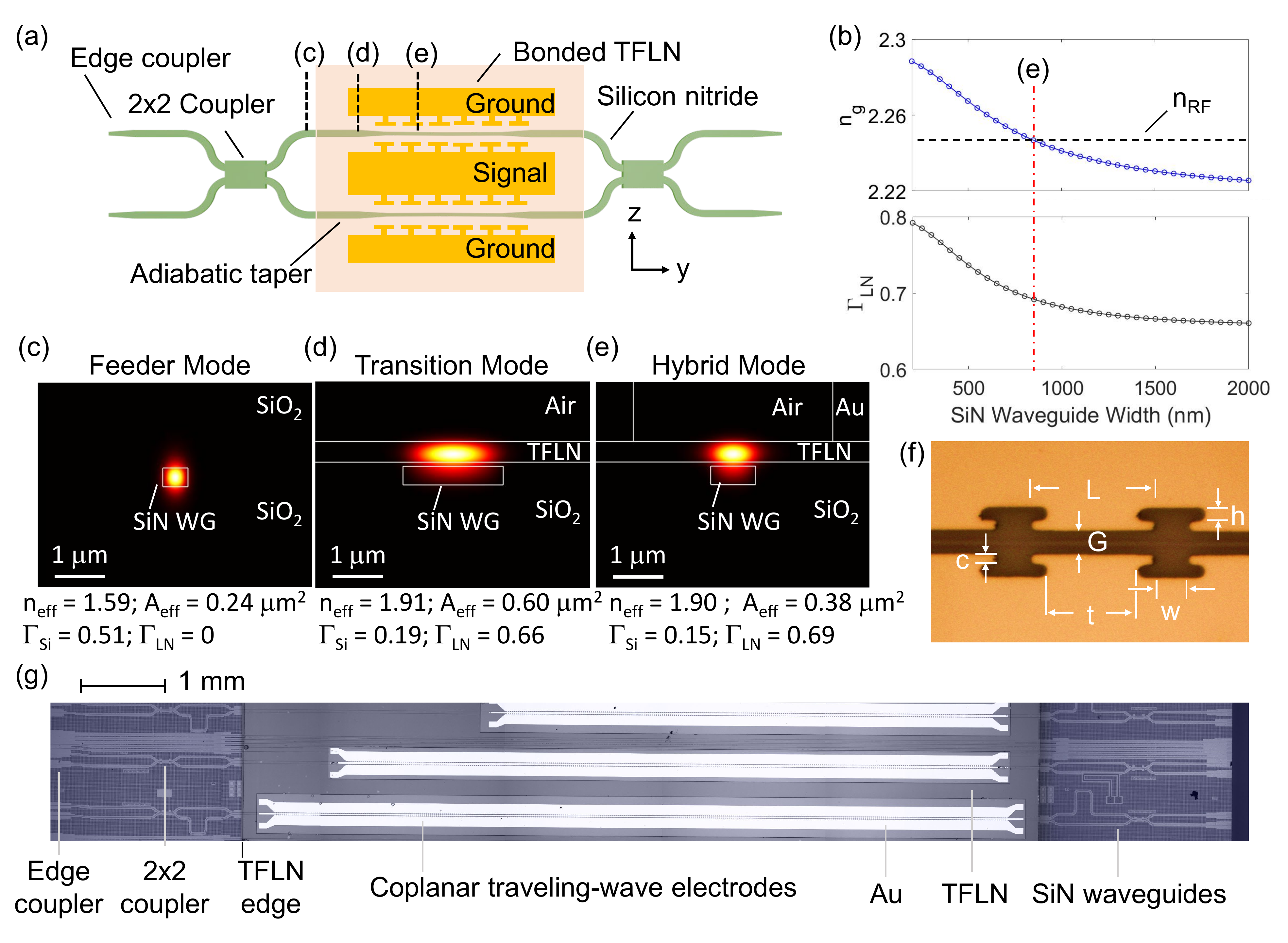}
\caption{(a) A top-view schematic of the hybrid bonded SiN-TFLN modulator (not to scale). (b) The simulated optical group index, $n_g$, and confinement integral of light in the TFLN region, $\Gamma_\mathrm{LN}$, as a function of the SiN waveguide width. The vertical dot-dashed red line corresponds to the mode shown in panel (e). The horizontal dashed black line corresponds to the simulated RF index at 110 GHz for the RF transmission line. The cross-sectional Poynting vector magnitude is shown of the (c) feeder SiN waveguide (no TFLN), (d) transition SiN waveguide, and (e) the phase-shifter section hybrid waveguide.  (f) A section of the RF transmission line with dimensions discussed in the text. (g) An optical microscope image of hybrid chip with a few different MZM structures.} 
\label{fig-concept}
\end{figure*}

\section{Hybrid MZM Concept and Design}
Figure \ref{fig-concept}(a) shows a top-down schematic of the hybrid MZM device. The optical index and mode fraction contained in TFLN can be varied by changing only the SiN width, as shown in Fig.~\ref{fig-concept}(b). The input and output sections are defined in SiN waveguides of height 0.18 $\mathrm{\mu m}$ and width 0.5 $\mathrm{\mu m}$ with an oxide cladding, whose TE-polarized fundamental optical mode profile is shown in Fig.~\ref{fig-concept}(c).  In contrast to hybrid silicon waveguides~\cite{weigelBondedThinFilm2018a,valdez110GHz1102022,valdez110GHz1102022}, the lower refractive index of SiN requires a different transition design. The feeder waveguide makes a transition before the edge of the bonded TFLN region to the mode profile shown in Fig.~\ref{fig-concept}(d), where the SiN width is increased to 2.0 $\mathrm{\mu m}$.  Once past the bonded edge, the waveguide is then adiabatically narrowed down over a length of 100 $\mathrm{\mu m}$ to a width of 0.9 $\mathrm{\mu m}$, which pushes most of the light into the TFLN region. This mode, shown as Fig.~\ref{fig-concept}(e), is used in the EO phase shift section with the x-cut TFLN. 

\begin{figure*}[tbh]
\centering
\includegraphics[width=\linewidth, clip=true]{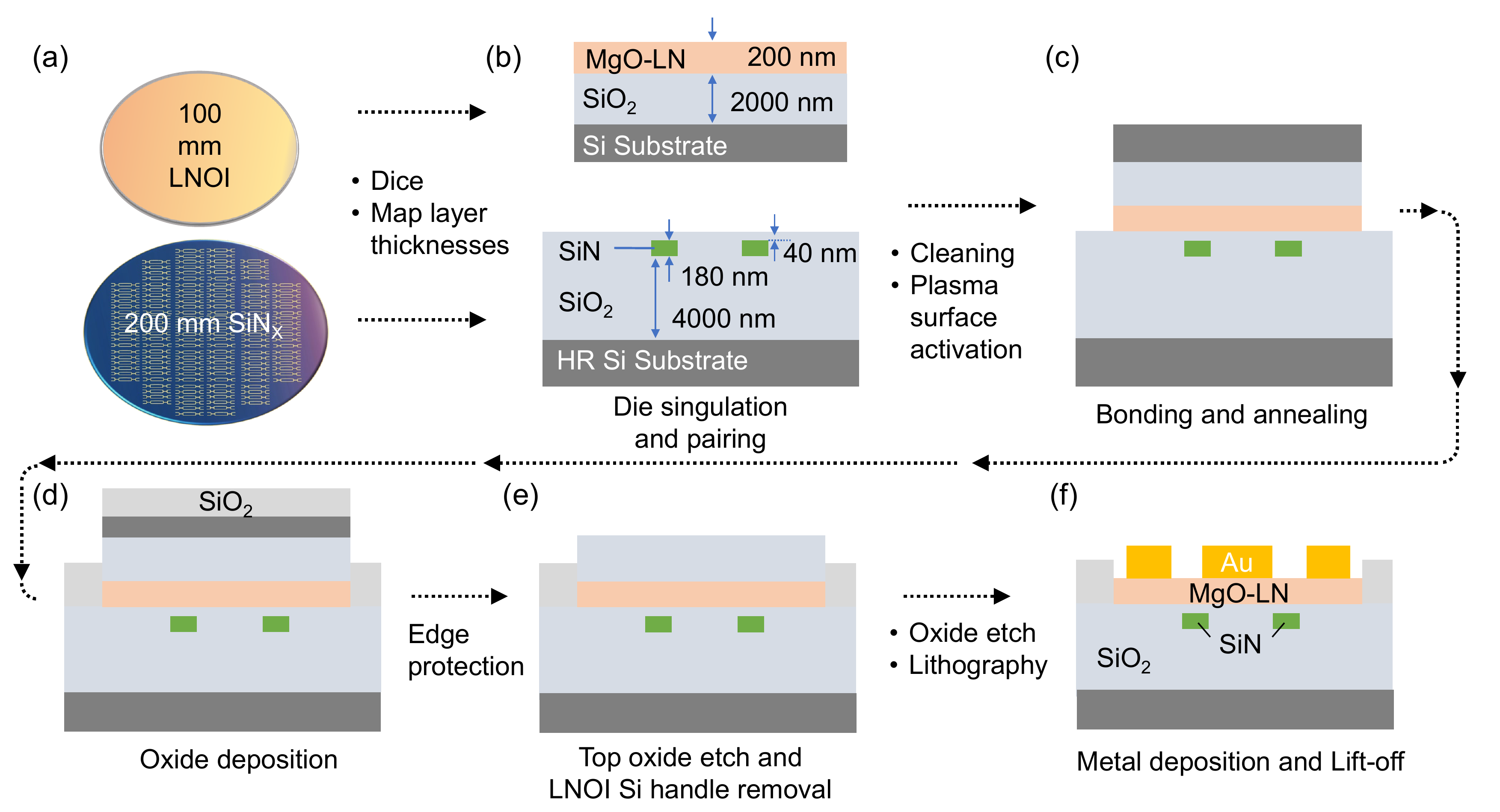}
\caption{Process flow diagram showing the main steps in the fabrication of the hybrid TFLN-SiN modulators. (Cross-sections are schematic diagrams, not to scale.)} 
\label{fig-fabrication}
\end{figure*}
The RF traveling-wave electrodes were designed using software (PathWave EM Design, Keysight Technologies) as coplanar ground-signal-ground traveling-wave structures. Slow-wave electrodes were used to match the optical and RF indices of refraction, using capacitively-loaded unit cells ~\cite{jaegerSlowwaveElectrodeUse1992,sakamotoNarrowGapCoplanar1995,rosaMicrowaveIndexEngineering2018,kharelBreakingVoltageBandwidth2021,chenHighPerformanceThinfilm2022}. A section of the slow-wave electrode structure is shown in Fig.~\ref{fig-concept}(f). Simulations were performed using the indicated parameters to adjust the RF index and dispersion. As the distance between the inner electrode edge and the inner T-rail edge increases [$h$ in Fig.~\ref{fig-concept}(f)], the RF wave becomes slower which results in a larger RF effective index. Furthermore, the T-rail stem width [$t$ in Fig.~\ref{fig-concept}(f)] also affects the RF wave velocity, with narrower $t$ corresponding to a slower RF field. The EOM device reported here used $G = 4\ \mathrm{\mu m}$, $h=2\ \mathrm{\mu m}$, $L = 20\ \mathrm{\mu m}$, and $t = 15\ \mathrm{\mu m}$.  The dashed lines in the top sub-panel of Fig.~\ref{fig-concept}(b) shows the simulated RF index ($n_\mathrm{RF} = 2.24$) which matches the optical group index of the mode shown in Fig.~\ref{fig-concept}(e).

Figure \ref{fig-concept}(g) is a stitched microscope image of a portion of one of the microchips, showing a few EOM devices fabricated along the east-west direction. The MZM shown in the middle of Fig.~\ref{fig-concept}(g) is the 0.8 cm long MZM reported here. The other devices on the chip share the same TFLN and SiN layer thicknesses, but have different SiN waveguide widths and coupler designs to address different wavelength bands. As such, many different devices can be fabricated in one bonding step. The electrode design includes flare-outs at the ends to facilitate landing of standard GSG probe tips. The SiN feeder waveguides are visible outside the bonded region, and transition seamlessly to the hybrid waveguides under the bonded region. The $2\times 2$ couplers at the input and output sections are designed as multi-mode interference (MMI) couplers using the feeder mode shown in Fig.~\ref{fig-concept}(c) and are 8 $\mathrm{\mu m}$ wide by 96 $\mathrm{\mu m}$ long to facilitate a 3-dB splitting ratio. While not affecting the design or performance of the EO phase-shift section, the overall insertion loss can be improved by using multiple SiN layers in the transition region~\cite{boyntonHeterogeneouslyIntegratedSilicon2020}. Also, the waveguide-fiber edge couplers were defined here as inverse tapers in a single SiN layer, and were not optimized for stand-alone performance.

\section{Fabrication Process}
Figure~\ref{fig-fabrication} shows an outline of the fabrication process in which hydrophilic direct bonding was used to integrate TFLN on top of SiN waveguides. The SiN waveguides were fabricated on a 200 mm wafer using a Si-photonics foundry process. After SiN lithography, $\text{SiO}_2$ was deposited using a plasma-enhanced chemical vapor deposition (PECVD) process, followed by a chemical mechanical planarization (CMP) process to achieve a planar and bondable surface. The oxide layer thickness was about 40~nm over the SiN waveguides, measured by ellipsometry on test sites. A smooth planarized surface was obtained with a root mean squared surface roughness less than 0.3~nm, measured by atomic-force microscopy. Lithium-Niobate-On-Insulator (LNOI) wafers (100~mm diameter) with 200~nm thickness 5 mol-\% magnesium oxide (MgO) co-doped x-cut LN on 2 $\mathrm{\mu m}$ thick buried oxide, and an approximately 500 $\mathrm{\mu m}$ thick Si handle were commercially procured (NanoLN, Jinan Jingzheng Electronics Co., Ltd.).  Both the LNOI and SiN wafers were singulated into individual dies and cleaned using Piranha and the RCA-1 process 1 (1:1:5 ratios of $\mathrm{NH}_4\mathrm{OH}$, $\mathrm{H}_2\mathrm{O}_2$ and $\mathrm{H}_2\mathrm{O}$)  to remove the organic and particle contaminants.  A plasma surface activation was performed using oxygen-based RF plasma (150 watts and 53 mbar, for 30 seconds), followed by a dip in ammonium hydroxide and de-ionized (DI) water to improve the hydrophilicity. 

The bonding was achieved by contacting the LN and planarized $\mathrm{SiO}_2$ surfaces at room temperature using a customized apparatus. This initial contact bonding is mediated by weak van der Waals forces. After room temperature bonding, the bonded sample was annealed using temperature cycles up to 300$^\circ \mathrm{C}$ under applied pressure. Ramp-up and ramp-down rates of 1$^\circ \mathrm{C}$/min were used during annealing. The annealing process converts the hydrogen bonds into covalent bonds at the interface, thereby improving the bond strength~\cite{xuGlassonLiNbO3HeterostructureFormed2018}. A 2 $\mathrm{\mu m}$ thick layer of $\mathrm{SiO}_2$ was deposited using PECVD to passivate the portion of the surface of the SiN chip which is outside the bonded TFLN region (mainly covering the edge couplers, 2x2 couplers and feeder waveguides). This was followed by a temporary polymer coating to protect the edges~\cite{mereModularFabricationProcess2022}. Hydrofluoric (HF) acid was used to etch the $\mathrm{SiO}_2$ that was deposited on top of the LNOI Si handle. Then, the Si handle was removed using xenon difluoride ($\mathrm{XeF}_2$) based dry etching. Lithography followed by oxide etching of the LNOI buried oxide layer was performed to access the surface of LN. A direct-writer laser lithography process (MLA 150, Heidelberg Instruments) was used to define and pattern the CPW electrode design. E-beam evaporation was used to deposit 20~nm thick titanium and 750~nm thick gold, followed by a lift-off process to complete the fabrication. 

\section{Measurements}
To measure the EO performance, a distributed feedback laser (784 nm wavelength; ThorLabs, Inc.) was coupled to the chip with lensed tapered fibers (Oz Optics). The on-chip optical power was about 0~dBm (1 mW). High-speed ground-signal-ground RF probes rated to 110 GHz (Picoprobe Model 110, GGB Industries) sourced and terminated the traveling-wave electrodes and were used with standard low-loss 1.0 mm RF cables and connectors. 

The hybrid mode, shown in Fig.~\ref{fig-concept}(e) is suitable for the straight phase-shift section, but does not support tight bends for microring resonators which can be used to estimate optical propagation loss. Therefore, the optical insertion loss (IL) of the MZM section, including the two $2\times 2$ couplers, transitions to the hybrid mode, and the phase shift sections, was estimated using the following subtractive procedure: first, the fiber-to-fiber loss was measured to be about 20 dB, which is primarily from the un-optimized edge coupling loss. Then, a test chip, fabricated using the same process on the same wafer, was measured which contained passive SiN waveguides of different lengths. From this cut-back method measurement, a propagation loss for the mode shown in Fig.~\ref{fig-concept}(b) was calculated as 1.6 dB/cm, and an edge coupling loss of 3.4 dB per facet. Taking into account the length of the feeder waveguides on the MZM chip (about 4.2 mm), the IL of the MZM was about 12 dB at 784 nm. Since we used a fixed wavelength laser for these measurements and the devices are not necessarily biased at the peak of the transmission, the minimum IL may be slightly lower (better) by about 1.5 dB based on the transmission versus voltage curves reported below.   

\begin{figure*}[tbh]
\centering
\includegraphics[width=\linewidth, clip=true]{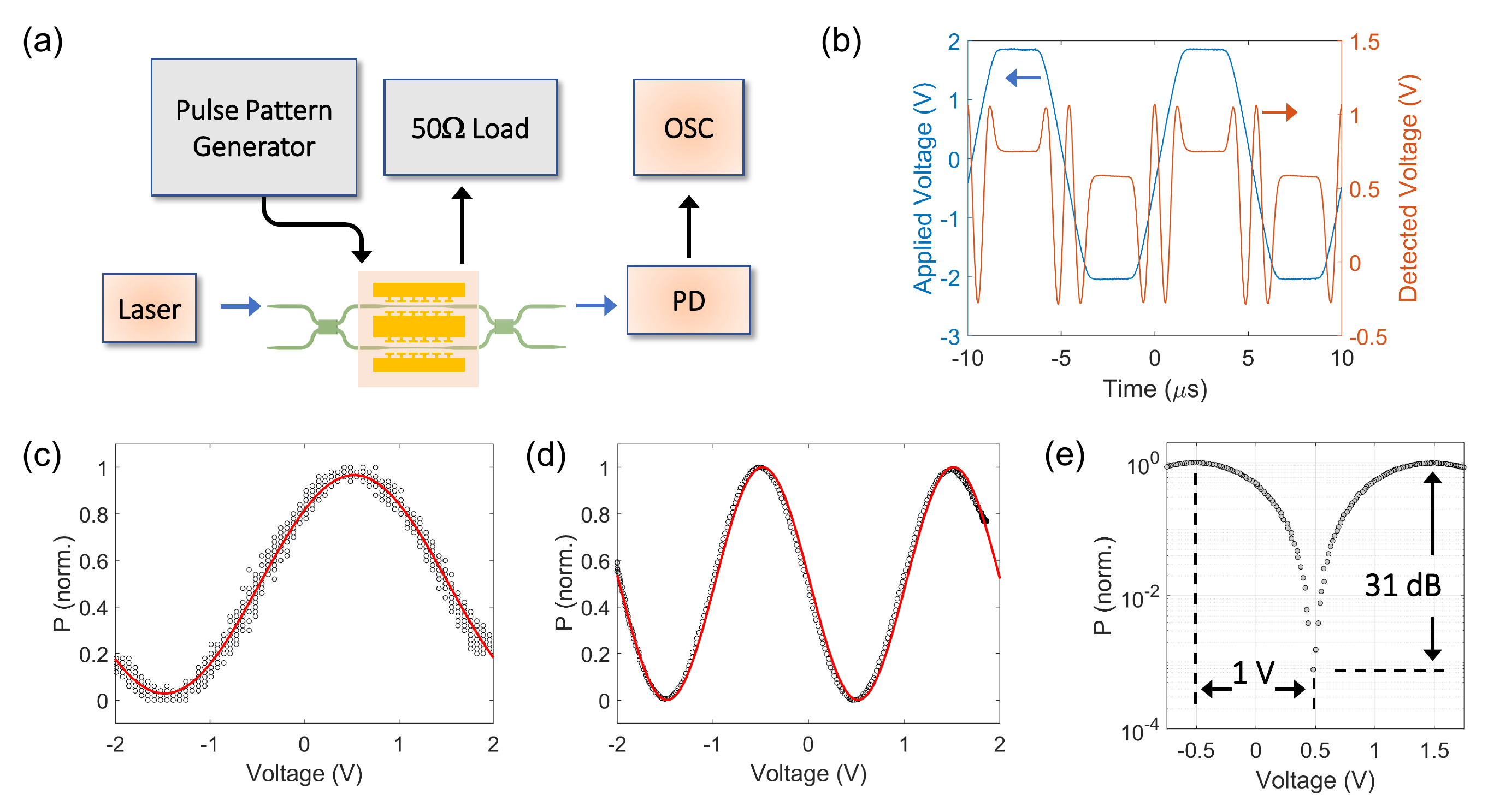}
\caption{(a) Schematic for measurement of the half-wave voltage, $V_\pi$. (b) Example of the (overdriven) optical response (red line) to the trapezoidal electrical drive signal (blue line). The vertical axis for the optical response is on an arbitrary linear scale determined by the photodetector. (c,d) The normalized optical power measured as a function of applied voltage for the hybrid MZM with length 0.4 cm [panel (c)] and 0.8 cm [panel (d)]. The solid lines are fits to the data using the squared-cosine functional form, which yields $V_\pi$. (e) A magnified portion of the data in panel (d) with a logarithmic vertical scale to quantify the extinction ratio.} 
\label{fig-VpiL}
\end{figure*}

Figure~\ref{fig-VpiL} describes the half-wave voltage ($\mathrm{V}_\pi$) measurement method and results. As shown in the schematic drawing, Fig.~\ref{fig-VpiL}(a), a 100 kHz trapezoidal signal generated by a pulse pattern generator (81110A, Agilent Technologies) was used to drive the modulator with 4 V peak-to-peak voltage. The modulated optical signal was detected with a photodetector (FD310-VIS-VIS, MenloSystems) and the electrical signal was recorded on an oscilloscope. An example of such a trace is shown in Fig.~\ref{fig-VpiL}(b). Because the applied voltage was greater than $V_\pi$, the over-drive can easily be seen in the optical oscilloscope trace. This waveform was post-processed using software to map the optical transmission to the applied voltage, and a cosine-squared fit was used to find $V_\pi$. Trapezoidal waveforms were used because, as shown in our previous work, they can more clearly show drift effects, if present, than triangular voltage ramps~\cite{mereModularFabricationProcess2022}. Figures~\ref{fig-VpiL}(c) and \ref{fig-VpiL}(d) show the modulated signals as a function of applied voltage, resulting in a measured $\mathrm{V}_\pi$ of 2.0 V and 1.0 V for the 0.4 cm and 0.8 cm long MZM devices, respectively ($V_\pi L =0.8$ V.cm). A portion of the trace shown in Fig.~\ref{fig-VpiL}(d) is shown in Fig.~\ref{fig-VpiL}(e) using a logarithmic scale for the vertical axis, from which an extinction ratio of about 31 dB was measured. This high degree of interferometric cancellation confirms that the couplers are well balanced and the mode transitions to the hybrid mode preserve the TE polarization state.  

Compared to the telecom wavelength regime, instrumentation to measure the high-speed EO response of modulators in the NIR wavelengths is limited. A calibration process is reported in Supplementary Information, Section~1. A combination of measurements was used to characterize the high-frequency EO response (EOR) from 1 to 110 GHz. (At short wavelengths, combinations of techniques have been used since the earliest reports of  modulators~\cite{gee17GHzBandwidth1983}.) Sinusoidal RF waveforms between 1 and 50 GHz were generated from an RF sweeper (83651B, Hewlett Packard, Inc.). From 1 GHz to 30 GHz, the modulated signals were measured with a high-bandwidth visible photodetector (Model 1471, Newport Corporation) and a high-speed sampling oscilloscope (DCA-X, Keysight Technologies). A fiber-pigtailed semiconductor optical amplifier (BOA785S, ThorLabs, Inc.) was used after the chip to amplify the optical waveform. The peak-to-peak voltage of the detected signal was recorded for each modulation frequency in steps of 1 GHz. For frequencies from 20 GHz onwards, an OSA (AQ6317B, Ando Electric Co., Ltd.) was used to measure the spectrum of the modulated signals. The EOR was determined by tracking the peaks of the carrier signal and the generated sidebands from modulation~\cite{shiHighspeedElectroopticModulator2003,leeBroadbandModulationLight2002,haffnerLowlossPlasmonassistedElectrooptic2018a,renaudSub1VoltHighBandwidth2022}.  From 20 GHz to 50 GHz, the sinusoidal modulation signal was generated directly by the RF sweeper. For 47 GHz to 78 GHz, an active multiplier was used (SFA-503753420-15SF-E1, Eravant), and also for 72 GHz to 110 GHz (SFA-753114616-10SF-E1, Eravant). For each of the four measurement bands, the amount of RF power delivered to the chip was calibrated (see Supplementary Information, Section 2). Measurements made in the overlapping RF frequency ranges show smooth continuity to within a small fraction of a dB.  

\begin{figure*}[tbh]
\centering
\includegraphics[width=\linewidth, clip=true]{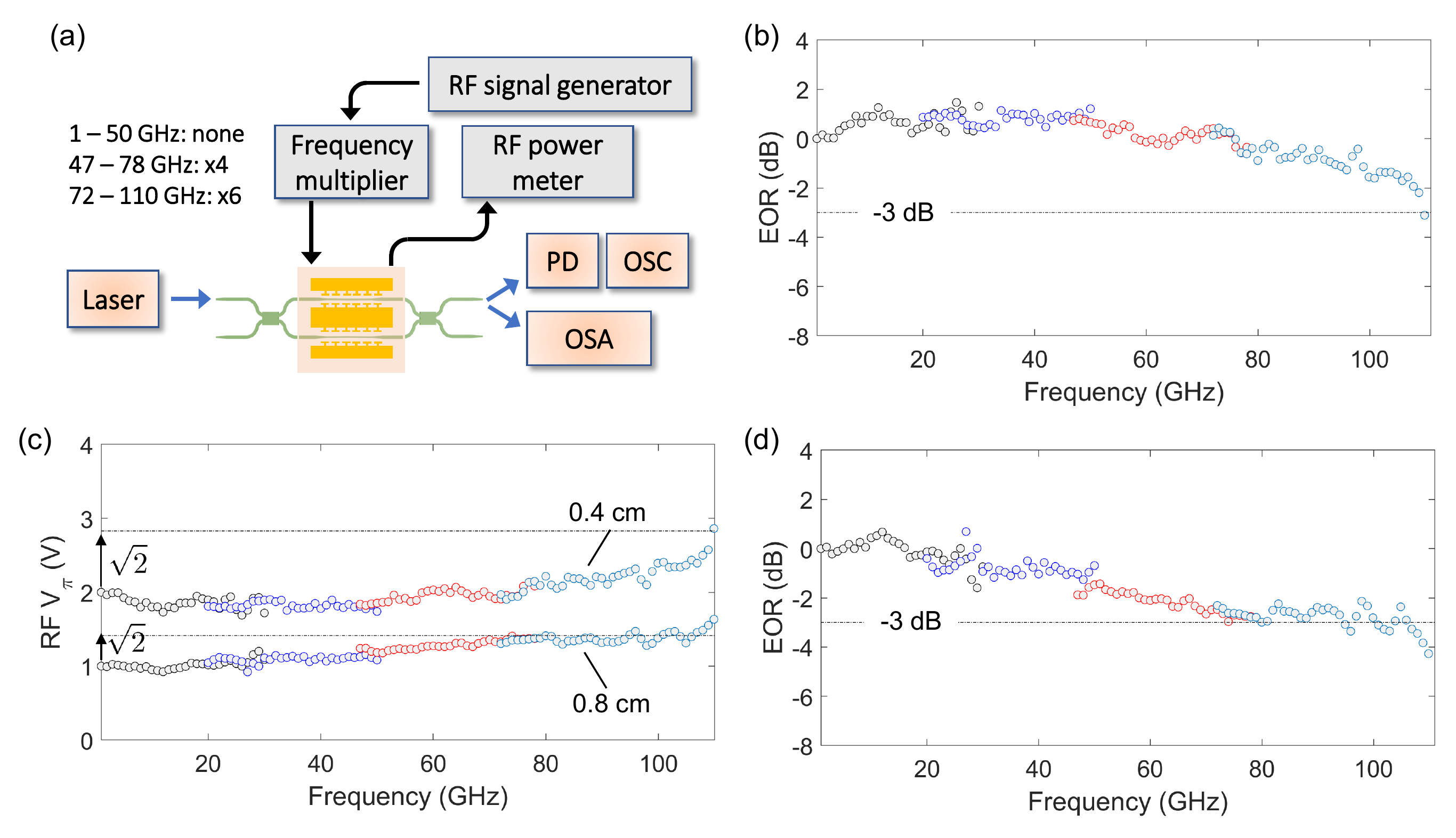}
\caption{(a) Schematic for measurement of the electro-optic response (EOR). (b,d) EOR measurement over 1 to 110 GHz of a hybrid TFLN-SiN MZM with phase-shift length 0.4 cm and 0.8 cm, respectively. The four colors for the data points represent the four RF measurement bands as described in the text. Each trace is normalized to the corresponding EOR at 1 GHz. (c) $V_\pi(f)$ normalized to 100 kHz for the two MZMs. The dashed lines show $\sqrt{2} V_\pi(\mathrm{DC})$ in each case.}
\label{fig-EORdata}
\end{figure*}

The EOR for two MZMs, with phase-shifter lengths 0.4 cm and 0.8 cm, are shown in Figs.~\ref{fig-EORdata}(a) and \ref{fig-EORdata}(b) respectively. The four colors in each curve represent the EOR measurements from the four overlapping RF frequency bands: 1 GHz to 30 GHz in black color, 20 to 50 GHz in blue, 47 GHz to 78 GHz in red, and 72 GHz to 110 GHz in purple. Generally, a small amount of noise was present at the last two or three points of each range where use the components slightly past their stated range. We have not removed or averaged out these artifacts for clarity, but they can usually be dropped in favor of the data points from the next band. The significance of the 3-dB frequency $f_0$ is that the EOR rolls off as $[1+(f/f_0)^2]^{-1}$ as $f$ increases above $f_0$. The EOR response does not roll off below 3 dB (shown as a dashed line) except at the far edge of the measured RF range, above 100 GHz, which we therefore take as the 3-dB roll-off frequency.    

\section{Discussion}
\subsection{Half-wave Voltage Length Product} 
The low-frequency half-wave voltage length product ($V_\pi L$) of a hybrid LN MZM is given by the following equation~\cite{weigelDesignHighbandwidthLowvoltage2021}
\begin{equation}
V_\pi L = \frac{1}{2} \frac{n_\mathrm{eff} \lambda G}{{n_e}^4 r_{33} \Gamma_\mathrm{mo}}
	\label{eq-VpiL}
\end{equation}
where $n_\mathrm{eff}$ is the effective refractive index of the optical mode, $\lambda$ is the wavelength, $G$ is the effective electrode gap distance between the ground and signal lines, $n_\mathrm{e}$ is the extraordinary index of refraction of LN, $r_{33}$ is the linear Pockels coefficient in the crystal z-direction along which the RF fields are oriented, and $\Gamma_\mathrm{mo}$ is the mode overlap integral between the optical mode and RF mode. $L$ is the length of the phase-shift section. The factor of 2 in the denominator is included as the structure is driven in a push-pull configuration. Compared to devices at telecom wavelengths, the NIR hybrid MZM achieves a smaller $V_\pi L=0.8$ V.cm because of three factors: the direct role of $\lambda$ in the numerator of Eq.~(\ref{eq-VpiL}), the higher refractive index $n_\mathrm{e}$ at shorter wavelengths, and a reduced electrode gap $G$ which is permitted by the tighter mode confinement without incurring high optical propagation loss from interacting with the metal electrodes. In previous work, we have reported $V_\pi L = 3.1$ V.cm for a hybrid MZM around $1.55\ \mathrm{\mu m}$~\cite{valdez110GHz1102022}; the reduced $V_\pi L = 0.8$ V.cm of this device implies a power reduction of about 12 dB may be achieved for electrical driver circuits at NIR wavelengths.   

\subsection{Frequency-dependent Half-wave Voltage}
The EOR can be reported as a frequency-dependent $\mathrm{V}_\pi$, given by the following equation~\cite{gopalakrishnanPerformanceModelingBroadband1994,howertonBroadbandTravelingWave2002}: 
\begin{equation} 
V_\pi(f) = V_\pi(\mathrm{DC}) \, 10^{-m(f)/20}
\end{equation}
where $m(f)$ is the measured EOR, and is shown in Fig.~\ref{fig-EORdata}(c) for the two MZMs with phase-shifter lengths 0.4 cm and 0.8 cm. $V_\pi$ (DC) is taken as the value of $V_\pi$ at 100 kHz measured for each MZM as described earlier. An increase of $V_\pi$(DC) by a factor of $\sqrt{2}$ corresponds to the EO $|S_{21}|$ of the modulator decreasing by 3 dB. $V_\pi(f)$ is relevant in calculating system requirements as the voltage needed for a $\pi$ phase shift generally increases at higher frequencies.  

\subsection{3-dB Bandwidth to Half-wave Voltage Ratio}
\begin{figure}[tbh]
\centering
\includegraphics[width=\linewidth, clip=true]{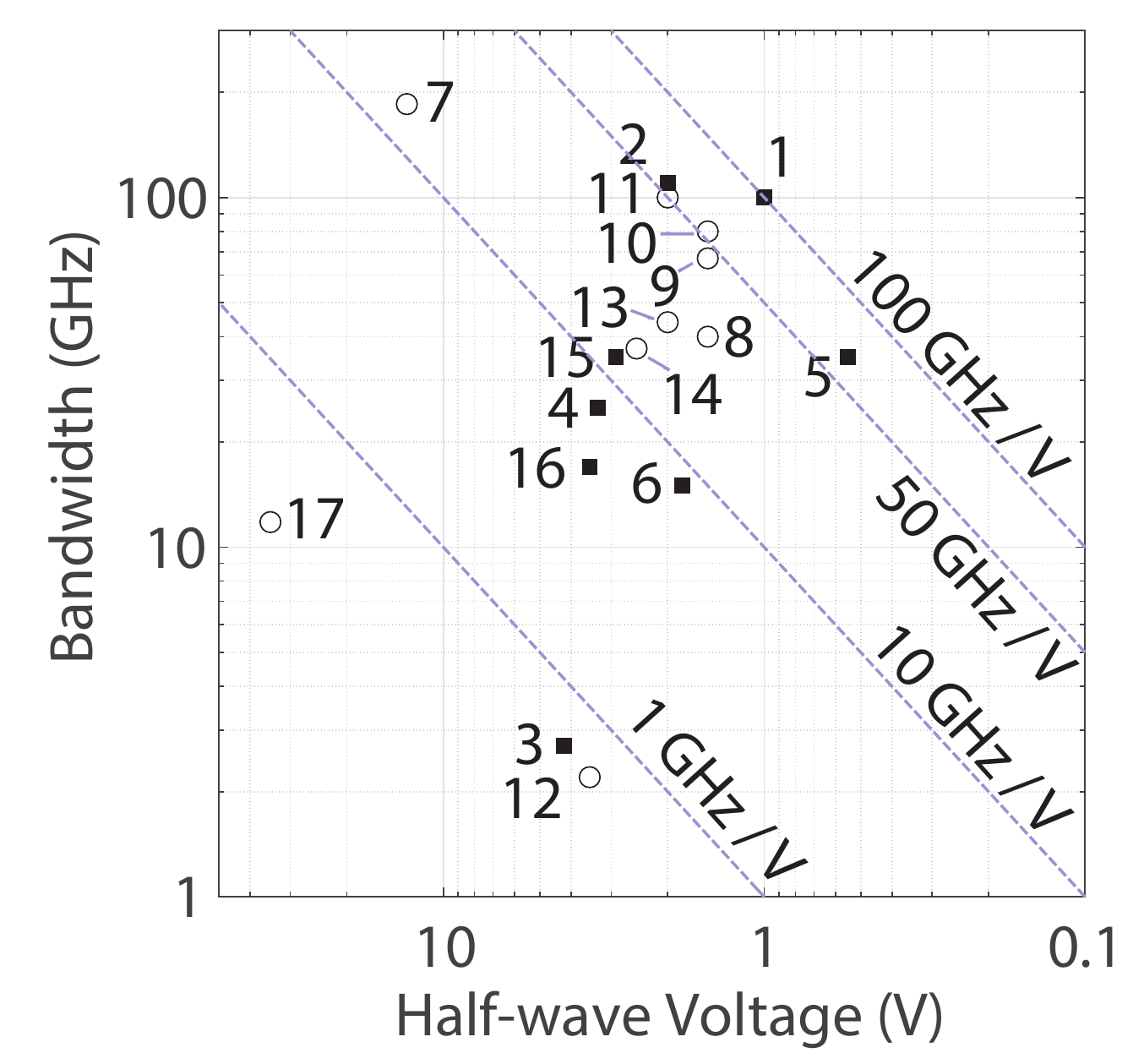}
\caption{Scatter plot of 3-dB EO Bandwidth (GHz) versus half-wave voltage (V) for various MZM devices with slope lines at 1, 10, 50 and 100 GHz/V indicated.  The correspondence between the numerical labels for the data points and the citation-list numbers is provided in the Supplementary Information, Section~\ref{sec-scatter}. Labels 1 and 2 are this work. Black squares represent short-wavelength modulators (wavelengths below $1\ \mathrm{\mu m}$) whereas open circles are at longer wavelengths (typically, 1310 nm or 1550 nm).} 
\label{fig-scatter}
\end{figure}

The hybrid MZM with $L=0.8$ cm is the first modulator, to our knowledge, that achieves 100 GHz EO bandwidth and $V_\pi = 1\ \mathrm{V}$ simultaneously. Figure~\ref{fig-scatter} shows a collection of reported MZM performance as a two-dimensional scatter plot with half-wave voltage on the horizontal axis (right to left) and 3-dB EO bandwidth on the vertical axis. In this representation, points to the upper-right (north-east) section are desirable, and dashed lines with various slope efficiencies are indicated.  

Resonant EO modulators, which use different design principles and operate in a different portion of the performance space, are not shown in Fig.~\ref{fig-scatter}. Among these, two notable EOMs are the graphene and plasmonic ring resonators~\cite{phareGrapheneElectroopticModulator2015,haffnerLowlossPlasmonassistedElectrooptic2018a}. The former uses only a monolayer of EO material and yet has shown an efficiency of about 3 GHz/V (30 GHz / 7.5 V driving voltage). The latter can achieve very high modulation bandwidths (possibly exceeding 1 THz), and at the measurement limit, an efficiency of about 33 GHz/V (110 GHz / 3.3 V).  On the other hand, a microresonator based device may exhibit spectral ripple from the cavity free-spectral range, is  sensitive to wavelength (or bias) and may have a smaller extinction ratio compared to MZM devices.   

In conclusion, we have demonstrated hybrid SiN-TFLN EO modulators with very high bandwidth and low half-wave voltage. No etching or patterning of the LN layer is required. Standard wafer-scale SiN manufacturing was used to define the optical routing, splitting, and tapering of the waveguides, and the top oxide layer of this wafer was prepared with a CMP process to provide a bondable surface to TFLN. The half-wave Voltage Length product ($V_\pi L$) was 0.8 V.cm at 784 nm. The RF EO response of the 0.4 cm long ($V_\pi=2\ \mathrm{V}$) and 0.8 cm long ($V_\pi=1\ \mathrm{V}$) MZM devices drops off past the 3 dB line (referenced to 1 GHz) only beyond 100 GHz. The achieved 3-dB Bandwidth to Half-wave Voltage Ratio of 100 GHz/V establishes a new benchmark in EO modulator performance. NIR photonics applications including communications and signal processing may now benefit from having as much EO bandwidth as the best telecom-band MZM's, but with an order-of-magnitude lower electrical power requirements. Scalability of NIR integrated photonics will be facilitated by the simple fabrication method of these hybrid MZM devices with today's silicon photonics platforms.  

\begin{acknowledgments}
The authors thank: R. Rouland (UC San Diego) for technical assistance, X. Wang, M. R\"using, P. O. Weigel and J. Zhao (formerly of UC San Diego) for earlier contributions and discussions; A. Lentine, N. Boynton, T. A. Friedman, S. Arterburn, C. Dallo, A. T. Pomerene, A. L. Starbuck, D. C. Trotter and A. Kodigala (Sandia National Laboratories, Applied Microphotonic Systems) for discussions and fabrication assistance; C. Coleman, R. Scott, G. Lee, B. Szafraniec, G. Vanwiggeren, V. Moskalenko, K.K. Abdelsalam and C. Langrock (Keysight Technologies) for discussions and measurement assistance. Part of this work was performed at the San Diego Nanotechnology Infrastructure (SDNI) of UCSD, a member of the National Nanotechnology Coordinated Infrastructure, which is supported by the National Science Foundation (Grant ECCS-2025752). This research was developed in part with funding from the Defense Advanced Research Projects Agency (DARPA) and the U.S. Government. This paper describes objective technical results and analysis. The views, opinions and/or findings expressed are those of the authors alone and should not be interpreted as representing the official views or policies of the Department of Defense or the U.S. Government
\end{acknowledgments}

%

\cleardoublepage
\centerline{\Large Supplementary Information}
\setcounter{section}{0}
\setcounter{equation}{0}
\setcounter{page}{1}
\setcounter{figure}{0}
\renewcommand{\thepage}{S\arabic{page}}
\renewcommand{\thesection}{S\arabic{section}}
\renewcommand{\theequation}{S\arabic{equation}}
\renewcommand{\thefigure}{S\arabic{figure}}
\section{LCA and OSA measurement}
\label{SI-section1}
\begin{figure*}[tbh]
\centering
\includegraphics[width=0.9\linewidth, clip=true]{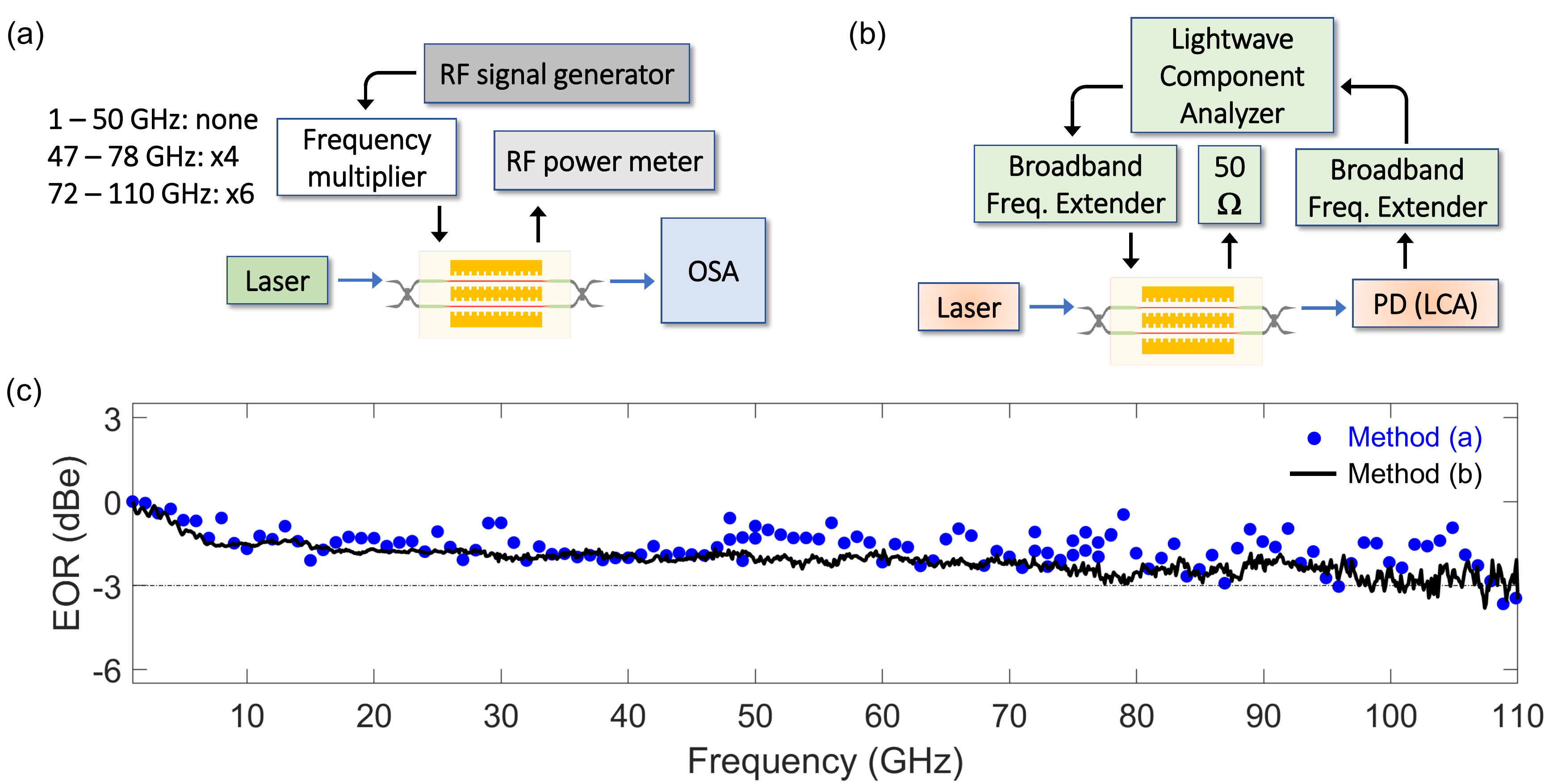}
\caption{(a) Schematic for measurement `method (a)' based on a sideband measurement on an optical spectrum analyzer, with the measurement performed in overlapping RF frequency bands, as indicated, and using RF multipliers (x4 and x6) to reach higher RF frequencies. (b) Schematic for `method (b)' using a single-connection Lightwave Component Analyzer (LCA) instrument which spans the entire frequency range, with built-in broadband frequency extenders. (c) Comparison of electro-optic response (EOR) measurements made using the two methods of the same chip, which is a hybrid TFLN-Si Mach-Zehnder modulator similar to the devices reported here.} 
\label{fig-EORCalib}
\end{figure*}

Calibration of the experimental apparatus was performed by a comparison between a measurement using a Lightwave Component Analyzer (LCA, N4372E, Keysight Technologies) and the sideband measurement on an optical spectrum analyzer. Similar experiments have been reported elsewhere, for example, Fig.~11 in Ref.~\cite{shiHighspeedElectroopticModulator2003} and were performed here for a batch of hybrid TFLN MZM devices. Figure~\ref{fig-EORCalib} shows measurements performed on a hybrid LN Mach-Zehnder modulator at 1550 nm~\cite{valdez110GHz1102022}. In this device, the underlying waveguide layer used Si rather than SiN but the change is not relevant for this discussion.  

The schematic for the sideband measurement is shown in Fig.~\ref{fig-EORCalib}(a), and for the LCA measurement in Fig.~\ref{fig-EORCalib}(b). Figure \ref{fig-EORCalib}(c) shows the EOR measured in these two ways over frequencies in the range of 1 to 110 GHz (referenced to 1 GHz in each case). Although more noise was present in the banded measurement performed using the OSA, the two methods agree in measuring the EOR trend over a wide frequency range, and predict the same roll-off and 3-dB EO bandwidth. This comparison allows us the sideband method to be used for measuring the EOR of the short-wavelength modulators, where there is no LCA instrument. The sideband measurements reported in Fig.~\ref{fig-EORdata} have less noise than in Fig.~\ref{fig-EORCalib} because of gradual improvements in our setup.

\section{Electro-optic Response Calculations}
We consider the optical spectrum of the output for a push-pull x-cut LN modulator with sinusoidal RF modulation, following the derivation in Ref.~\cite{shiHighspeedElectroopticModulator2003}. For an input light field amplitude $E_i$ at optical frequency $\omega_0$, and ignoring loss, the output field is 
\begin{gather}
E_o = \frac{E_i}{2} \, \exp(j \omega_0t) \big[ \exp(j m_1 \, \cos \omega_m t) \\
+ \exp(j m_2 \, \cos \omega_m t + j\phi) \big]
\end{gather}
where $\omega_m$ is the external sinusoidal modulation signal frequency, $m_1$ and $m_2$ are the phase modulation amplitudes in the two interferometer arms, and $\phi$ is a constant phase difference between the two interferometer arms. For push-pull operation, $m_1 = -m_2 = m$ and 
\begin{equation}
m = \frac{\pi}{2} \frac{V}{V_\pi}
	\label{eq-def-m}
\end{equation} where $V$ is the applied voltage and $V_\pi$ is the half-wave voltage.

By expanding the output in a series of Bessel functions, we can write the intensity of the $k$-th frequency component as
\begin{equation}
I(\omega_0 + k \omega_m) = \frac{1}{2} I_i \, {J_k}^2(m) \left[ 1+ (-1)^k \cos \phi\right]
\end{equation} 
where $I_i$ is the input intensity. 

\begin{figure}[h]
\centering
\includegraphics[width=\linewidth, clip=true]{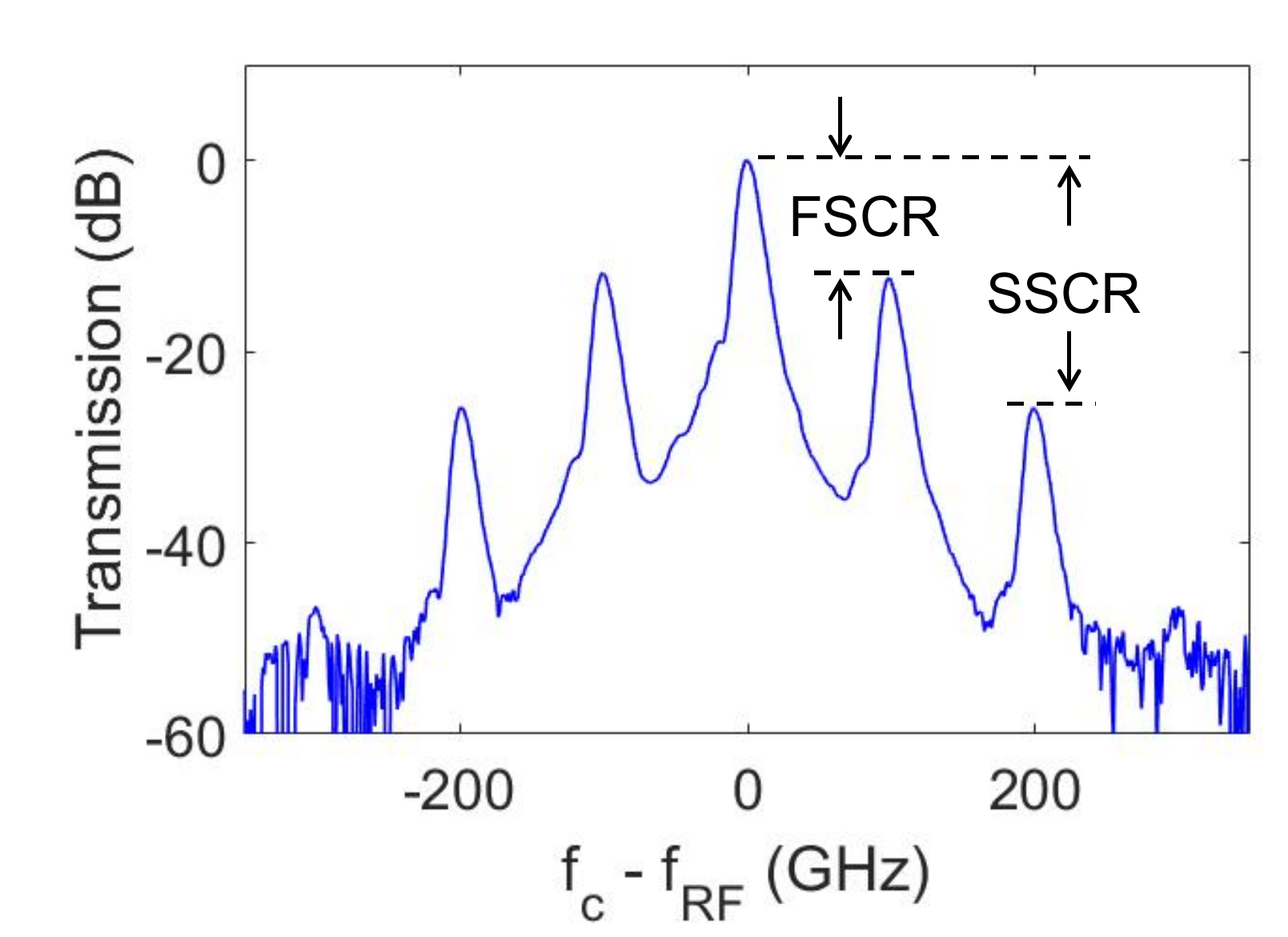}
\caption{An example of the first-sideband to carrier ratio (FCSR), described by Eq.~(\ref{eq-ratio1}), and the second-sideband to carrier ratio (SSCR), described by Eq.~(\ref{eq-ratio2}), for RF modulation at 100~GHz.} 
\label{fig-S-SCR}
\end{figure}

From the optical spectrum captured at each sinusoidal RF modulation frequency, the two ratios, first-sideband to carrier ratio (FSCR), $I(\omega_0+ 1\omega_m)/I_i$, and second-sideband to carrier ratio (SSCR), $I(\omega_0+ 2\omega_m)/I_i$, are determined. From these two measured quantities, the unknown parameters $m$ and $\cos \phi$ can be inferred, using the equations 
\begin{eqnarray}
\frac{I(\omega_0+ \omega_m)}{I_i(\omega_0)} &=& {\left(\frac{{J_1}(m)}{{J_0}(m)}\right)}^2\frac{(1- \cos \phi)}{(1+ \cos \phi)}  \label{eq-ratio1}\\
\frac{I(\omega_0+ 2\omega_m)}{I_i(\omega_0)} &=& {\left(\frac{{J_2}(m)}{{J_0}(m)}\right)}^2 \label{eq-ratio2}
\end{eqnarray}
The FSCR and SSCR are shown in a representative example in Fig.~(\ref{fig-S-SCR}). The linewidths are limited by the OSA resolution, but the peak values can be accurately determined as they are well above the noise floor. 

Note that often, the small-signal modulation limit is assumed (i.e.,~$m\ll 1$), so that $J_1(m) \approx m/2$ and $J_0(m) \approx 1$, which makes the inversion algebraically simple. Here, because $V_\pi$ is expected to be small and $m$ is therefore not small compared to unity, we do not make this approximation, and directly solve for the pair of equations shown above using numerical root-finding. 

Once $m$ is determined at each RF modulation frequency $f$, we use Eq.~(\ref{eq-def-m}) to solve for $V_\pi(f)$. This is accomplished by writing Eq.~(\ref{eq-def-m}) as 
\begin{equation}
\begin{split}
20\ \log_{10} m(f) & = 20\ \log_{10} \frac{\pi}{2} + 20\ \log_{10} V(f) \\
& - 20\ \log_{10} V_\pi(f).
\end{split}
	\label{eq-def-m-log}
\end{equation}

Also, based on Eq.~(2) in the main text,
\begin{equation}
\mathrm{EOR}(f) = 20\ \log_{10} V_\pi(\mathrm{DC}) - 20\ \log_{10} V_\pi (f)
\end{equation}
and Eq.~(\ref{eq-def-m-log}) can be used to write
\begin{equation}
\begin{split}
\mathrm{EOR}(f) &= 20\ \log_{10} V_\pi(\mathrm{DC}) - 20\ \log_{10} \frac{\pi}{2} \\
&+20\ \log_{10} m(f) - 20\ \log_{10} V(f).
\end{split}
\label{eq-def-m-log-2}
\end{equation}

The EOR curve is referenced to its value at a low frequency and thus, the frequency-independent terms on the first line of Eq.~(\ref{eq-def-m-log-2}) can be dropped when calculating how EOR rolls off with $f$. Two terms remain, the first of which, $20\ \log_{10} m$, is found from the optical spectrum as described above. The second term, $20\ \log_{10} V$, is linearly proportional to the RF power delivered to the MZM, labeled $P_\mathrm{chip}$. (In our earlier work, we have reported VNA-based measurements of similar electrode structures which show that the impedance $Z$ of the MZM electrode structure is approximately constant over this frequency range.) 

$P_\mathrm{chip}(f)$ can be inferred as follows: a measurement is performed without the MZM chip, but under the same source settings to record the RF power generated by the source (either the RF signal generator directly, or with the 4x or 6x multipliers in place), $P_\mathrm{src}(f)$, followed by the input-side cable and connectors whose loss is labeled $\mathrm{IL}_\mathrm{cab}(f)$. The power measurements were performed using a calibrated thermal RF power meter (NRP110T.02, Rohde \& Schwarz). We also account for the loss of the GSG probe as provided by the manufacturer data sheet, labeled $\mathrm{IL}_\mathrm{GSG}(f)$. (Both the IL quantities are written as positive-valued numbers; a higher number indicates greater loss.) Then, the RF power delivered to the chip, $P_\mathrm{chip} = P_\mathrm{src}-\mathrm{IL}_\mathrm{cab}-\mathrm{IL}_\mathrm{GSG}$. 

Thus, the EOR curve can be found by calculating $20\ \log_{10} m$ and subtracting $P_\mathrm{chip}$ (units: dBm). The ``DC'' value (more accurately, at 100 kHz) is obtained from a separate measurement using trapezoidal waveforms as described in the main text and is re-introduced to calculate the $V_\pi(f)$ curve.  

As stated in the main text, this spectral-domain method is used for those RF frequencies where the first and second sidebands are clearly distinguishable from the carrier peak, which is for RF frequencies above 20 GHz in our experimental setup. The relative EO response at lower RF frequencies can be measured using a time-domain method, similar to that used in the measurement of $V_\pi L$, in which the modulated waveform is captured on a sampling oscilloscope with a high-bandwidth (about 35 GHz) photoreceiver and the peak-to-peak magnitude is recorded. The range of frequencies over which the two methods overlap is used to vertically shift and match the time-domain response (which is scaled by the responsivity of the photoreceiver and optical amplifier, if used) to the spectral-domain EO response curve, which is self-calibrated [see Eq.~(\ref{eq-ratio1}) and (\ref{eq-ratio2})] since the carrier and sidebands can be clearly distinguished. This is accomplished by calculating a mean value for both traces in the overlapping frequency range, and vertically offsetting the time-domain data to the spectral-domain curve.    

\section{References Listed in Fig.~\ref{fig-scatter}}
\label{sec-scatter}
The data points in Fig.~\ref{fig-scatter} are labeled with numbers, and these labels correspond to the following numbered references in the unified citation list.  
\begin{table}[h]
\begin{tabular}{|c|c|} \hline
Figure label & Reference number \\ \hline
1 & This work (L=0.8 cm) \\ 
2 & This work (L=0.4 cm) \\ 
3 & \cite{celikHighbandwidthCMOSvoltagelevelElectrooptic2022a} \\ 
4 & \cite{liHighModulationEfficiency2022a} \\ 
5 & \cite{renaudSub1VoltHighBandwidth2022} \\ 
6 & \cite{rogersNanosecondPulseShaping2016} \\ 
7 & \cite{leeBroadbandModulationLight2002} \\ 
8 & \cite{kieningerUltraHighElectroOpticActivity2017} \\ 
9 & \cite{ogiso67GHzBandwidth2017} \\ 
10 & \cite{yamazakiIMDDTransmissionNet2019} \\ 
11 & \cite{ozolins100GHzExternally2017} \\ 
12 & \cite{hirakiHeterogeneouslyIntegratedIII2017} \\ 
13 & \cite{langeLowSwitchingVoltage2016} \\ 
14 & \cite{prosykHighPerformance40GHz2012} \\ 
15 & \cite{walkerOptimizedGalliumArsenide2013} \\ 
16 & \cite{gee17GHzBandwidth1983} \\ 
17 & \cite{walkerHighspeedElectroopticModulation1987} \\ \hline
\end{tabular}
\end{table}

\end{document}